\documentclass[trackchanges]{aastex701}
\usepackage{float}
\usepackage{placeins}
\accepted{October 21, 2025}

\begin{document}

\title{Unveiling the Mixing and Transport Processes of Solar Wind and Planetary Ions in the Magnetopause Boundary Layer}

\author{Zhongwei Yang}
\affiliation{State Key Laboratoty of Solar Activity and Space Weather, National Space Science Center, Chinese Academy of Sciences, Beijing, China}
\email[show]{zwyang@swl.ac.cn} 

\author{Can Huang} 
\affiliation{Key Laboratory of Deep Petroleum Intelligent Exploration and Development, Institute of Geology and Geophysics, Chinese Academy of Sciences, Beijing, China}
\email[show]{huangcan@mail.iggcas.ac.cn}

\author{Xiaocheng Guo}
\affiliation{State Key Laboratoty of Solar Activity and Space Weather, National Space Science Center, Chinese Academy of Sciences, Beijing, China}
\affiliation{University of Chinese Academy of Sciences, CAS, Beijing, China}
\email[]{xcguo@swl.ac.cn}

\author{Riku Jarvinen}
\affiliation{Finnish Meteorological Institute, Helsinki, Finland}
\email[]{riku.jarvinen@fmi.fi}

\author{Binbin Tang}
\affiliation{State Key Laboratoty of Solar Activity and Space Weather, National Space Science Center, Chinese Academy of Sciences, Beijing, China}
\email[]{bbtang@swl.ac.cn}

\author{Wence Jiang}
\affiliation{State Key Laboratoty of Solar Activity and Space Weather, National Space Science Center, Chinese Academy of Sciences, Beijing, China}
\email[]{jiangwence@swl.ac.cn}

\author{Hui Li}
\affiliation{State Key Laboratoty of Solar Activity and Space Weather, National Space Science Center, Chinese Academy of Sciences, Beijing, China}
\affiliation{University of Chinese Academy of Sciences, CAS, Beijing, China}
\email[]{hli@swl.ac.cn}

\author{Chi Wang}
\affiliation{State Key Laboratoty of Solar Activity and Space Weather, National Space Science Center, Chinese Academy of Sciences, Beijing, China}
\affiliation{University of Chinese Academy of Sciences, CAS, Beijing, China}
\email[]{cw@swl.ac.cn}


\begin{abstract}

Kelvin-Helmholtz (KH) vortices are widely observed in astrophysics and heliophysics, including at Jovian and terrestrial magnetopauses, the Martian sheath-ionosphere boundary, the heliopause, and within stellar accretion disks. These vortices play a critical role in transporting mass, momentum, and energy across boundary layers. Magnetized planets such as Earth exhibit a higher incidence of fully rolled-up, nonlinear KH vortices compared to non-magnetized planets like Mars. In contrast to previous magnetohydrodynamic (MHD) studies, this work adopts a kinetic point of view to quantify ion mixing rates using three-dimensional global hybrid simulations, with Earth as a representative case. This approach enables automated identification of the KH-modulated,  corrugated magnetopause. For the first time, we provide a quantitative assessment of how solar wind conditions control {\bf solar wind} entry and subsequent mixing with magnetospheric ions via KH waves. We find that under northward interplanetary magnetic field (IMF) conditions, the flux of particles crossing the dayside magnetopause increases with solar wind dynamic pressure and peaks in the KH region. Notably, the KH-modulated low-latitude boundary layer thins as the dynamic pressure increases. Under southward IMF conditions, coupled reconnection and KH structures further enhance solar wind injection and boost magnetospheric ion escape in the dayside, especially near the subsolar point where reconnection intensifies this exchange. These results also shed light on the evolution of space environments and mass transport at magnetized planets in the heliosphere and beyond.

\end{abstract}



\section{Introduction} \label{sec:s1}

The Kelvin-Helmholtz instability (KHI), named after Lord Kelvin and Hermann von Helmholtz, arises in fluids experiencing shear either due to a velocity gradient within a continuous medium or a differential velocity at the boundary layer between two distinct fluid layers \citep{Chandrasekhar1961,Miura1982,Pu1983}. This phenomenon is widely observed in solar and heliophysics, manifesting in structures such as Earth's magnetopause \citep{Farrugia1998,Hasegawa2004}, Jupiter's rotating atmosphere \citep{Vasavada1998} and magnetosphere \citep{Montgomery2023}, the heliopause and inner heliosheath \citep{Borovikov2008,Borovikov2014,Fraternale2024}, interplanetary coronal mass ejections \citep{Nykyri2024a}, and in the plasma dynamics of the Sun and interstellar medium \citep[e.g.,][]{Foullon2011,Ofman2011,Berne2012}. These ubiquitous and fundamental KH waves play a critical role in the transport of energy and mass across boundary layers spanning a broad range of spatial and temporal scales. The extent of KH vortex evolution varies across different planetary environments.

Within our solar system, weakly magnetized planets such as Mars and Venus, which possess only crustal magnetic fields, allow their ionospheres and exospheres to interact directly with the solar wind \citep{Penz2004,Mostl2011,Poh2021}. Mars, for example, has observed a few highly curled KH vortex events \citep{Ruhunusiri2016,Zhang2018,Kim2021}, a rarity partly due to the ionosphere's density significantly surpassing the solar wind particle density, resulting in a mismatch that diminishes the growth rate of KH instabilities. The well-known unstable condition for the KHI is written as follows \citep{Chandrasekhar1961}.
\begin{equation}\label{eq1}
\gamma_{KH}^2=\frac{\rho_1\rho_2}{(\rho_1+\rho_2)^2}[\mathbf{k}\cdot(\mathbf{U_1}-\mathbf{U_2})]^2-\frac{1}{\mu_0(\rho_1+\rho_2)}[(\mathbf{k}\cdot\mathbf{B_1})^2+(\mathbf{k}\cdot\mathbf{B_2})^2]>0,
\end{equation}
where $\rho_i$, $\mathbf{U}_i$ and $\mathbf{B}_i$ ($i = 1, 2$) are the mass density bulk velocity, and magnetic field at each side across the boundary, respectively. Additionally, Mars's smaller size means that the contact region between the magnetosheath, ionosphere, or crustal fields is substantially smaller than that of those of magnetized planets with robust magnetospheres. In this case, the solar wind–ionosphere interaction on Mars is fundamentally different from that on magnetized planets. KH waves may facilitate the escape of ionospheric and atmospheric particles, potentially affecting the planet's habitability over geological timescales \citep[e.g.,][]{Penz2004,Lammer2013,Wang2023}. In contrast to Mars and Venus, magnetized planets possess extended magnetospheres, forming a boundary layer where the material density on both sides is comparably matched. This configuration makes it easier to observe fully rolled-up KH vortices, such as those seen in Earth's magnetopause \citep{Hasegawa2004}.

Taking Earth as an example of a magnetized planet, magnetic reconnection at low latitudes along the dayside magnetopause plays a key role in the transfer of mass, momentum, and energy under southward interplanetary magnetic field (IMF) conditions \citep{Dungey1961}. In contrast to southward IMF conditions, several mechanisms for the transport of solar wind plasma and energy across the magnetopause under northward IMF have been proposed based on in situ observations \citep{Taylor2008,Foullon2008,LinD2014,Plaschke2016}. One possible scenario is the KHI, driven by velocity shear across both sides of the magnetopause \citep{Miura1984,Song1992,Fairfield2000,Hasegawa2004}. The onset of the KHI typically occurs near the subsolar point and evolves along the flanks into developed KH vortex structures that propagate toward the magnetotail. This evolution provides ample time for KH waves in Earth's magnetopause to mature, frequently resulting in the observation of fully rolled-up KH vortices \citep{Hasegawa2017}. ARTEMIS observations show that KH waves near Earth (within 20 Earth radii) typically have periods of 100-300 seconds and wavelengths of 1-10 Earth radii, while those farther out (around the lunar orbit at 60 Earth radii) exhibit longer periods of 200-700 seconds and wavelengths of approximately 20-25 Earth radii \citep{Zhou2022}. It is worth noting that Mercury, possessing a weaker Hermean magnetosphere compared to Earth's dipole, also exhibits KH phenomena along its magnetospheric flanks, with KH waves crossing the satellite's trajectory on timescales of approximately 10 seconds \citep{Paral2013}. In contrast, Jupiter, a giant planet with a vastly larger rotational magnetosphere than Earth, presents a scenario where internal dynamics predominate over solar wind conditions in KH activity. The periodicity of observed KH waves observed at the Jovian magnetopause is generally on the order of hours \citep{Delamere2021}.

The excitation and evolution of the KHI along the magnetopause under various solar wind conditions have also been studied using local MHD and kinetic simulations \citep[e.g.,][]{Nykyri2001,Nakamura2004,Cowee2010,Ma2014a,Ma2014b,Blasl2022,Blasl2023}, global MHD simulations \citep[e.g.,][]{Raeder2003,Guo2010,Hwang2011,Hwang2012}, and global hybrid simulations \citep{Hwang2023,Yang2025}. Although the K-H instability has been suggested to play a critical role in atmospheric ion loss on Mars, the knowledge about its formation and evolution is still poor, due to the limitations of spacecraft missions and a dearth of dedicated simulation codes. Thus, \citet{Wang2023} conducted quantitative studies of particle injection and escape processes at non-magnetized planets like Mars using global hybrid simulations for the first time. Their simulation results showed good agreement with MAVEN observations. However, the solar wind entry and planetary particle loss at KH-modulated, corrugated magnetopauses of magnetized planets have not yet been investigated quantitatively using global hybrid simulations.

Furthermore, high-resolution observations from the MMS satellite revealed that KHI can occur under southward IMF conditions, potentially accompanied by magnetic reconnection events \citep[e.g.,][]{Eriksson2016,Li2023,Nykyri2024b}. Local 3D Full PIC simulations \citep{Nakamura2014,Nakamura2017} show that a reconnection-KH combined structure triggered by a certain southward IMF component can lead to a mass transfer rate nearly one order of magnitude higher than previous expectations for the KHI. This highlights the need for a global kinetic simulation-based statistical study of particle entry and escape across the magnetopause under varying IMF orientations and solar wind dynamic pressures.

In this paper, Earth is selected as a representative case among the myriad magnetized planets to investigate the impact of KH waves on the permeation of solar wind particles and the loss of planetary particles across the corrugated magnetopause under diverse solar wind conditions. The primary objective of this work is to investigate for the first time how KH-active regions are distributed across the magnetopause under different solar wind dynamic pressures and IMF orientations, and how these structures quantitatively affect global solar wind entry and magnetospheric particle escape.

\section{Simulation model} \label{sec:s2}

To answer the above questions, the magnetized planet-solar wind interaction has been simulated by a three-dimensional (3-D) global hybrid simulation platform RHybrid \citep{Jarvinen2018}. As in previous studies for magnetized planets Mercury and Earth \citep{Jarvinen2020,Jarvinen2024,Yang2024}, the model setup includes unperturbed, upstream solar wind ions injected into the simulation from the right boundary along the $-x$ direction with a drifting Maxwellian velocity distribution. Within the simulation domain, ion velocity distributions evolve according to model calculations that are self-consistently coupled with the evolution of the magnetic field. The initial state of the simulation employs a mirror-dipole magnetic field profile, commonly used in global MHD \citep{Raeder2003,Guo2010} and hybrid \citep{Lin2014} simulations for magnetized planets. Electrons are modeled as a massless charge-neutralizing fluid. The inner boundary is assumed at the geocentric distance of $r = 1.5~R_\mathrm{E}$ and is implemented as a perfectly conducting sphere on which precipitated particles are absorbed.

Usually, the simulated Earth radius is rescaled to a magnitude of approximately $10~d_\mathrm{i0}$ (the upstream solar wind ion inertial length) in order to save considerable computational costs in particle simulations and to ensure the appearance of an Earth-like magnetosphere \citep[e.g.,][]{Blanco-Cano2006,Lin2014,Palmroth2023}. In this paper, the rescaled Earth radius $R_\mathrm{E}^\prime$ is set to $1000~\mathrm{km}$. Uniform grid cells with a size of $\Delta\mathrm{x} = \Delta\mathrm{y} = \Delta\mathrm{z} = 0.1~R_\mathrm{E}$ are used throughout the simulation domain. The mesh grid dimensions in this study are set to $n_\mathrm{x}\times n_\mathrm{y}\times n_\mathrm{z} = 300\times 500\times 500$. A total of about five billion particles are used, and a typical time step is $\Delta\mathrm{t} = 0.01~\mathrm{s}$. The variation range of solar wind dynamic pressure across different cases spans from less than $0.5~\mathrm{nPa}$ to over $10~\mathrm{nPa}$; the IMF clock angle varies from $0^{\circ}$ to $360^{\circ}$, with cases spaced at intervals of $45^{\circ}$. More details of the basic parameter setup and case lists are summarized in Tables 1 and 2, respectively.

\section{Simulation results} \label{sec:s3}

\subsection{A reference case: northward IMF} \label{sec:s3.1}

\subsubsection{Nonlinear Kelvin-Helmholtz vortices in a global hybrid simulation} \label{sec:s3.1.1}
Figure 1 provides an overview of the simulation result for run 3 (a reference case, see Table 2) under a pure northward IMF condition, depicting the Earth's magnetosphere reaching a mature state (e.g., at $t=1020~\mathrm{s}$) under the continuous influence of a  $450~\mathrm{km/s}$ solar wind. Figure 1a shows that the KH wave structures in the low-latitude boundary layer (LLBL) originate near the subsolar point and gradually evolve into a nonlinear phase with fully rolled-up vortices as they propagate toward the nightside \citep[e.g.,][]{Kavosi2015}. In Figure 1b, the ion bulk velocity component $V_x$ exhibits distinct vortex characteristics in the KH region. In this hybrid simulation, we assign a unique identification (ID) to each particle, enabling us to track ions originating from both the solar wind and the magnetosphere separately. Solar wind ions are continuously injected from the upstream boundary (Figure 1c); magnetospheric ions originate from the mirror dipole region and remain captured within the mature magnetosphere after its formation (Figure 1d). It is observable that solar wind ions infiltrate the cusp region at high latitudes, while at lower latitudes, they permeate into the flanks of the magnetopause (MP) through the advanced evolution of KH waves. The aforementioned KH physical pictures under northward IMF conditions are consistent with previous MHD simulations \citep{Guo2010}, and numerous in situ observations from spacecraft such as Geotail \citep[e.g.,][]{Fairfield2000}, Cluster \citep[e.g.,][]{Hasegawa2004}, THEMIS \citep[e.g.,][]{Henry2017} and MMS \citep[e.g.,][]{Rice2022}. In the following sections 3.2 and 3.3, the impact of varying solar wind dynamic pressure and IMF clock angle on the ion flux entering and exiting the dayside magnetopause will be quantitatively investigated, respectively.

\subsubsection{Calculations of the mixing rate} \label{sec:s3.1.2}

To distinctly delineate the composition on either side of the KH boundary layer, at least two approaches for particle simulations are available: (1) the mixing fraction $\mathit{F}$ \citep{Daughton2014,Nakamura2021}, and (2) the mixing rate/occupation rate $\mathit{MR}$ \citep{Matsumoto2006}. The expression for the first is as follows
\begin{equation}\label{eq2}
\mathit{F}=(N_\mathrm{ms}-N_\mathrm{sw})/(N_\mathrm{ms}+N_\mathrm{sw}),
\end{equation}
where $\mathit{F}$ will vary continuously from $\mathit{F}=1$ in regions of pure magnetospheric plasma to $\mathit{F}=-1$ in regions of pure magnetosheath plasma 2. The subscripts ``$\mathrm{ms}$" and ``$\mathrm{sw}$" refer to the same species of particles originating from the magnetosphere and magnetosheath compressed solar wind, respectively. This method is helpful for analyzing the mutual permeation of materials across the boundary layer. The expression for the second method is defined below
\begin{equation}\label{eq3}
\mathit{MR}=-2|N_\mathrm{sw}/N_\mathrm{tot}-0.5|+1,
\end{equation}
where $N_\mathrm{tot}=N_\mathrm{sw}+N_\mathrm{ms}$. The two measures are equivalent (i.e., $\mathit{MR}=1-|\mathit{F}|$), and in principle either approach could be adopted. The advantage of using $\mathit{MR}$ is that it maps the result to [0,\,1], producing a clear peak in visualization at the boundary layer where mixing is strongest. Therefore, to automatically outline a boundary layer position for each case, we adopt this method in this paper.

For the northward IMF case above, the KH instability is most pronounced in low-latitude regions, as the instability criterion in equation (1) is typically satisfied where $\mathbf{k}\cdot\mathbf{B}$ is close to zero. Therefore, we use a cross-section of the magnetic equatorial plane from the 3-D global simulation to illustrate the computed mixing rate $\mathit{MR}$ in Figure 2 (the corresponding 3D results are shown in Figure 5). The regions of magnetosheath and magnetosphere on either side of the boundary transition to white. The colored regions represent the mixing of materials on both sides, with red indicating the peaks of the mixing rate, which outline the position of the magnetopause boundary modulated by nonlinear Kelvin-Helmholtz effects. The complete three-dimensional mixing rate profile for this northward IMF case will be presented and compared to the other cases in Section 3.3 (Figure 5). In summary, through the aforementioned calculations, we can obtain the envelope of the magnetopause.

\subsubsection{Calculations of the ion transport} \label{sec:s3.1.3}

Based on the previously obtained magnetopause envelope, we initially triangulate it \citep{Moreland2013}. Firstly, for the triangulated three-dimensional surface of the magnetopause, we can determine the area $\Delta\mathit{S}$ and normal vectors $\mathbf{\hat{n}}$ of each finite element. Secondly, based on particle data from the global hybrid simulation, we can track particles and calculate various moment information such as density, bulk velocity for both solar wind ions and magnetospheric ions. Then, the flux density of solar wind ions across the magnetopause can be calculated using $N_\mathrm{sw}\mathbf{V}_\mathrm{sw}\cdot\mathbf{\hat{n}}$ on each triangulated element. Using a similar calculation, the flux density of magnetospheric ions across the magnetopause can be determined using $N_\mathrm{ms}\mathbf{V}_\mathrm{ms}\cdot\mathbf{\hat{n}}$. In this paper, we focus on the solar wind ions entering the magnetosphere and the magnetospheric ions escaping. As shown in Figure 3c (run 3), the flux density of solar wind ions entering the magnetosphere is represented in blue (i.e., the negative part of the flux density), while the escaping magnetospheric ions are depicted in red (i.e., the positive part of the flux density). In summary, using the methods outlined above, we can perform quantitative statistical analyses of the flux density of ions entering and escaping the magnetopause under various solar wind conditions. For a more detailed description of the method, refer to \citet{Yang2025}.

\subsection{The impact of solar wind dynamic pressure} \label{sec:s3.2}

This section will conduct a statistical analysis of the impact of varying solar wind dynamic pressures on particle transport at the magnetopause. For runs $1\text{--}5$ in Table 2, we first calculate the ion mixing rate $\mathit{MR}$ at the magnetopause under different solar wind dynamic pressure conditions. As shown in Figure 2, from left to right, the nonlinear KH structures at the magnetopause vary for different solar wind dynamic pressures from $P_\mathrm{d}<0.5~\mathrm{nPa}$ to $P_\mathrm{d}>10~\mathrm{nPa}$. The findings indicate that under conditions of lower solar wind dynamic pressure, the magnetospheric boundary layer is not severely compressed, remaining relatively relaxed. This broader boundary layer leads to larger KH wavelengths. As demonstrated by laboratory experiments \citep{Thorpe1971} and corroborated by theory \citep{Hazel1972}, KH billows typically manifest wavelengths approximately 12$\text{--}$15 times the thickness of the boundary layer before instability onset. {\bf By analyzing the zoomed-in solar wind density gradient of the boundary-layer flank at earlier times, we find that the density boundary becomes sharper (i.e., the layer becomes thinner) under higher solar wind dynamic pressure conditions, resulting in smaller KH vortex scales.} For instance, on the dawn side ($Y<0$), the visible KH wave structures increase from approximately 3 under the $1.2~\mathrm{nPa}$ condition (Figure 2c) to $4\text{--}5$ under $3.4~\mathrm{nPa}$ (Figure 2d). Furthermore, fully rolled-up nonlinear KH structures are often accompanied by higher-order vortices and partial fragmentation. Figure 2e presents the magnetopause in an extremely compressed state under a high dynamic pressure ($10.7~\mathrm{nPa}$), where KH excitation becomes relatively slow, with only faint KH signals visible on the nightside ($X<0$). Previous theoretical and numerical studies have shown that compressible effects play an important role by gradually stabilizing the KHI when the flow becomes supermagnetosonic \citep{Miura1982}, without achieving complete stabilization because of the finite width of the velocity-shear layer \citep{Miura1992}. Based on these previous studies, in the case of high solar wind speed associated with high dynamic pressure (Figure 2e), the shear free energy may be converted into compressional waves and dissipated as heat within the boundary layer, rather than feeding the KHI.

In Figure 2, we have displayed the profiles of the magnetopause mixing rate under varying solar wind dynamic pressure conditions. For each case study, the three-dimensional distribution of both the solar wind ion entry flux density and the magnetospheric ion escape flux density on the derived magnetopause envelopes is shown in Figure 3. The colorbar range remains consistent across all case studies. It is worth mentioning that although the KH scales are larger in the low dynamic pressure cases, the flux density of ions entering and exiting at the magnetopause does not necessarily increase. Overall, as the solar wind dynamic pressure increases, there is a noticeable rise in ion flux density, particularly in low-latitude regions. We find that the primary region for solar wind ions entering the magnetosphere is the leading edge of the KH, facing the sun (in red). Conversely, escaping magnetospheric ions are elevated out of the magnetopause through the trailing edge of KH structures. They are then picked up and carried away by the solar wind in the magnetosheath, completing their escape process. Up to this point, we have gained a basic qualitative understanding of the conditions under different solar wind dynamic pressures. Next, we will quantitatively investigate the fluxes of entering solar wind and escaping magnetospheric ions in different sectors of the dayside magnetopause under varying solar wind dynamic pressure conditions.

Figure 4a consists of a group of concentric circles, with each circle from the inside out representing one of five case studies of increasing solar wind dynamic pressure (runs $1\text{--}5$). For each case study, the magnetopause is divided into six sectors along the dawn-dusk and north-south axes, each covering a $60^\circ$ clock angle. Within each sector, the average solar wind injection flux, denoted by $|flux|~\mathrm{(cm^{-3}km/s)}$, is quantified and labeled on the corresponding sector of the concentric circle for that case study. The flux unit used here directly combines the commonly used density unit $\mathrm{cm^{-3}}$ with the plasma bulk velocity unit $\mathrm{km/s}$ for an intuitive understanding. The statistical results from Figure 4a indicate that the average flux of entering solar wind ions generally increases across all sectors with rising solar wind dynamic pressure. Especially, the highlighted yellow areas represent the locations of the largest solar wind injection flux, which are observed specifically in the two sectors: dawn and dusk of the KH-modulated LLBL. Figure 4b employs the same visualization method to present the quantitative statistical results for escaping magnetospheric ions. The results show that ion loss in the low-latitude KH boundary layer increases with rising dynamic pressure, following a trend similar to that of the solar wind injection in Figure 4a. Furthermore, it is evident that the flux of solar wind entry exceeds that of planetary ion loss by approximately an order of magnitude. Figure 4 presents the statistical results at the final simulation time ($t=1020~\mathrm{s}$). Using the same approach, we have thoroughly examined the ion fluxes entering and leaving the magnetopause at other moments for all cases (see Appendix A). In summary, the conclusions drawn from these additional time points are consistent with the results presented above.

\subsection{The impact of IMF clock angle} \label{sec:s3.3}

In this section, we investigate the impact of the IMF clock angle on solar wind injection and planetary ion loss under the same solar wind dynamic pressure conditions. Figure 5 depicts the three-dimensional global distribution of the mixing rate under different IMF clock angle conditions (runs 3, 6–12 in Table 2). The viewpoint shown in Figure 5 is from the direction of the Sun towards Earth. Starting from the top and rotating clockwise, the eight case studies represent the IMF conditions changing from purely northward, to duskward, to purely southward, to dawnward, and finally returning to purely northward, with each case spaced by a $45^{\circ}$ clock angle interval. More detailed parameter settings are provided in Table 2. The red pointers in the eight black clocks in the middle part of Figure 5 indicate the directions of the IMF clock angles. The mixing rate calculation process and the color scale are the same as those used in Figure 2. The top position, displaying the complete three-dimensional distribution of the mixing rate from the northward reference case discussed in the previous section. First, when the IMF $B_\mathrm{z}$ component is northward, the KH-modulated magnetopause exhibits a distinct ``orange segment" shape. The areas with a higher mixing rate, indicated by warmer colors, are predominantly found on the dayside low-latitude flanks of the magnetopause. Second, when the IMF has a $B_\mathrm{y}$ component, the magnetopause exhibits an ``S" shape \citep{Trattner2007}. The regions of KHI migrate away from the equator, exhibiting a dawn-dusk asymmetry. Then, as the IMF $B_\mathrm{z}$ component is southward, the magnetopause undergoes magnetic reconnection. \citet{Nykyri2024b} highlights that KH-induced reconnection has become a hot topic in recent MMS observational studies. However, electron-scale processes exceed the scope of this paper. Local Full PIC simulations \citep{Nakamura2008,Fermo2012,Hasegawa2017} are most effective for studying reconnection triggers associated with KH vortices, whereas the global hybrid simulations used in this work are primarily employed to examine their global impacts \citep{Wang2023,Hwang2023}. Although global Full PIC simulations for solar-terrestrial interactions were previously developed \citep{Yang2016}, a complete 3D study remains infeasible due to CPU resource limitations and is beyond the scope of this paper.

Similar to Figure 3, Figure 6 visualizes the ion fluxes entering and exiting the magnetopause, outlined by the peak mixing rates in Figure 5. Let us begin at the top and examine each case study in a clockwise direction. The first case (run 3), featuring a northward IMF, is shown at the top and has been detailed in Figure 3 as the reference case. The flux of particles across the magnetopause peaks in regions of maximum KH growth rates, which are broad and confined to the equator, away from the subsolar point, consistent with previous MHD simulation results by F98 \citep{Farrugia1998} and Cluster observations \citep{Foullon2008}. For the northeast IMF case (run 6)—the second in our clock-angle sequence—we observe that the particle inflow and outflow flux peaks become narrower. They migrate away from the equator, shifting southward on one flank and northward on the other, depending on the sign of the clock angle. This behavior is consistent with KHI growth rates predicted by F98 on the dayside \citep{Foullon2008}. Conversely, the KH structures on the upper right and lower left sides of this case study are relatively weaker. The third case (run 7) on the right features a duskward IMF, where the entire particle ingress and egress region exhibits an ``S" shape. This ``S" shaped magnetopause configuration was previously observed in early MHD simulations of magnetized planets like Earth and Jupiter \citep{Sibeck2014, Sarkango2019}. The dawn-dusk oriented IMF quasi-perpendicularly pressures the upper left and lower right areas of the magnetopause, causing the ``S"-shaped ends to extend toward the higher latitudes near the cusp. In the fourth case (run 8), the IMF is southeast-oriented. On one hand, due to the presence of a southward IMF component, magnetic reconnection begins to appear atop the magnetopause. On the other hand, the direction of the IMF draping places the high-latitude parts of the ``S" shape in the upper right and lower left of the magnetopause. Compared to cases with a northward $B_z$ component in the IMF, the fluxes of ions entering and exiting in this southward-leaning case become somewhat chaotic due to reconnection modulation, yet the ``orange segment" KH ingress and egress patterns are still discernible. At the bottom, the purely southward case (run 9) is shown, where the presence of reconnection boosts the ion fluxes entering and leaving the magnetopause. A quantitative analysis of these flux variations will be provided in Figure 7. The next three cases on the dawn side (runs $10\text{--}12$), essentially mirror the dawn cases (runs $8\text{--}6$), and will not be reiterated here. In summary: (1) When the northward IMF component is present, the ion fluxes in the mid- to low-latitude KH-dominated regions of the dayside magnetopause are the most significant. The KH activity region and the peaks of ion fluxes at the magnetopause migrate with the rotation of the IMF clock angle; (2) When the By IMF component is present, an ``S" shaped distribution of the KH and particle exchange regions may occur, primarily due to the sheath magnetic field lines wrapped around the magnetopause; (3) When the southward IMF component triggers magnetic reconnection, forming a mixing structure with KH, it effectively boosts solar wind injection and planetary particle loss.

Finally, we conducted a quantitative statistical analysis of the cases under different clock angles. Unlike Figure 4, which used concentric circles to demonstrate the effects of dynamic pressure, here in Figure 7 we employ the same visualization method as in Figures 5 and 6 to illustrate the impact of clock angle. As shown in Figure 7a, eight circular diagrams arranged clockwise represent the cases for different clock angles, with the IMF direction indicated by the red needle in the small black clock at the center. Each circle (case study) corresponds to a magnetopause divided into six sectors (similar to the approach in Figure 4). The top case features a northward IMF, with the sector values already described in Figure 4; here, it can be summarized simply as the magnetic equator being higher than the polar regions. Due to significant variations in flux across different sectors under various clock angles, the average flux values in Figure 7 are presented in logarithmic form. The statistical results indicate that when the IMF has a southward component, more solar wind ions enter the magnetopause and even migrate towards the poles. The corresponding physical processes involve KH mixing with magnetic reconnection. However, whether the IMF is northward or southward, the flux near the magnetic equator is always the greatest. Correspondingly, Figure 7b displays the average flux statistics for escaping magnetospheric ions under different IMF clock angle conditions. The overall results are similar to the trend of solar wind ion injection, with higher fluxes at the equator than at the poles, and an enhanced particle transport associated with a southward component of the IMF. Under southward IMF conditions, there is a slight asymmetry between the dawn and dusk sides. This could be explained by the local magnetopause being more nonstationary, with magnetic reconnection occurring at random locations.

\section{Conclusions and discussions} \label{sec:s4}

This work presents 3-D global hybrid simulations of the Earth's magnetosphere in different solar wind pressure and IMF clock angle conditions. We conducted a qualitative and quantitative study on the global distribution and sub-sector statistics of entering and escaping ion fluxes at the magnetopause modulated by fully rolled-up nonlinear Kelvin-Helmholtz waves, identified through the mixing rate method. The main conclusions are:

(1) Hybrid simulations allow us to individually track ions from both the solar wind and the interior of the magnetosphere. Leveraging this capability, we computed the mixing rate across the entire simulation area. Utilizing the peak values of this rate, we can automatically delineate a three-dimensional contour of the magnetopause for each case study at every instance. The method for calculating the mixing rate, originally developed for local PIC simulations, has been expanded to global hybrid simulations, and we have demonstrated its effectiveness.

(2) By studying the mixing rate and ion flux across the magnetopause under different solar wind dynamic pressure conditions, we observed that under northward IMF conditions, the mass exchange primarily occurs near the magnetic equator, where KH waves are most easily excited or reach their fullest development. The fluxes of entering solar wind ions and escaping magnetospheric ions are strongest near the magnetic equator, and their average intensity increases with increasing solar wind dynamic pressure. Additionally, we found that under extremely high dynamic pressure conditions (e.g., $>10~\mathrm{nPa}$), the magnetopause is significantly compressed, the boundary layer thins, and the corresponding scale of KH waves decreases.

(3) Under identical solar wind dynamic pressure conditions, we also studied the impact of the IMF clock angle on the mixing rate and ion flux. The results indicate that under southward IMF conditions, the flux of ions entering and exiting the magnetopause is greater than that under northward conditions, especially near the subsolar point where reconnection intensifies this exchange. For magnetized planets like Earth, the flux of solar wind ions entering the dayside usually exceeds the flux of escaping magnetospheric ions by approximately an order of magnitude.

This paper pioneers the use of particle simulation techniques to investigate material exchanges at the magnetopause, with findings applicable under conditions where solar wind remains stable over minute-long scales. In cases where the solar wind contains complex structures like current sheets, TDs, or RDs, as seen in scenarios \citep[e.g.,][]{Guo2021,Wang2021,Lin2022}, a tailored approach for each situation is necessary. Looking ahead, it may be possible to expand the computational domain to include the dawn-dusk side foreshock under Parker IMF conditions to explore the characteristics of KH structures. Additionally, the influence of various factors on KH, such as Earth's tilt angle changing with seasons as it orbits the sun \citep{Kavosi2023}, should also be considered.

In this study, we do not need to assume that the boundary is stationary for the calculation to work, because the mixing rate method allows us to obtain the boundary at any given time and shape. However, since boundary motion is not explicitly considered, this method may not accurately reflect the actual solar wind entry flux. Relying solely on the bulk velocity of solar wind particles may lead to inaccuracies, particularly when the magnetopause motion is comparable to the solar wind velocity. Although previous global simulations have also neglected magnetopause motion \citep[e.g.,][]{Palmroth2003,Jing2014}, \citet{Brenner2021,Brenner2023,Brenner2025} have begun to include its effect on energy transport in storm-time MHD simulations. Due to computational constraints, all current global hybrid simulations still employ scaling (e.g., $R_\mathrm{E}\sim10~\mathrm{d_i}$ in \citet{Lin2014}; $R_\mathrm{E}\sim28~\mathrm{d_i}$ in \citet{Palmroth2023}; $R_\mathrm{E}\sim10~\mathrm{d_i}$ in \citet{Omelchenko2024}; and $R_\mathrm{E}\sim12~\mathrm{d_i}$ in this work), and thus MHD simulations remain the most suitable tool for accurately investigating the influence of magnetopause motion in realistic scales. With the advancement of computational capabilities, there is also hope to further extend this mixing rate method to the interstellar-medium–modulated heliopause, and to incorporate motion corrections during rapid boundary-shift events in the future.

\begin{acknowledgments}
Authors are grateful to Shahab Fatemi from Swedish Institute of Space Physics and Xueyi Wang from Auburn University for helpful discussions. This work was supported by the National Natural Science Foundation of China (42274210, 42188101, and 42150105), the National Key R\&D Program of China (2021YFA0718600), and the Specialized Research Fund for State Key Laboratories of Solar Activity and Space Weather. C.H. is supported by the Key Research Program of the Chinese Academy of Sciences (KGFZD-145-2023-15). R.J. received funding from the European Research Council (Grant agreement No. 101124960). Simulations were performed using the RHybrid code distributed under the open source GPL v3 license by the Finnish Meteorological Institute (github.com/fmihpc/rhybrid).
\end{acknowledgments}

\appendix

\section{Flux over time}
Figure 4 above presents the solar wind injection and planetary particle loss fluxes statistically obtained at the end of the simulations ($t=1020~\mathrm{s}$). In Figure 8, we provide additional statistical information over time. The solid curves represent the average solar wind ion injection fluxes derived from the dawn and dusk sectors at low latitudes. The curves, with increasing thickness, represent results from runs $1\text{--}5$ (see Table 2) under varying solar wind dynamic pressure conditions. Particularly for the reference case (run 3), we trace the flux evolution back to approximately $t=700~\mathrm{s}$, when the magnetosphere had just reached its mature state, i.e., the positions and sizes of the magnetopause and bow shock became relatively stable. The magnetosphere reaches maturity at slightly different times for each case. Therefore, we primarily focus on the final mature stage ($t>850~\mathrm{t}$). The corresponding results for planetary ion loss are denoted by dashed curves. In summary, results obtained at different times in Figure 8 are quantitatively consistent with the conclusions drawn at a fixed time in Figure 4.

\begin{deluxetable*}{cccccccccccc}
\tablenum{1}
\tablecaption{Global hybrid model setup and solar wind conditions for the reference case with a northward IMF (i.e., run 3 in Table 2).\label{tab:tab1}}
\tablehead{Parameters & Value}
\startdata
Number of grid cells $(n_\mathrm{x}\times n_\mathrm{y}\times n_\mathrm{z})$ & $300\times500\times500$\\
Grid cell size $(\Delta\mathrm{x})$ & $100~\mathrm{km}=R_\mathrm{E}^\prime/10$\\
Time step $(\Delta t)$ & $10~\mathrm{ms}$\\
SW bulk velocity vector $[V_\mathrm{x},\ V_\mathrm{y},\ V_\mathrm{z}]$ & [-450, 0, 0]$~\mathrm{km/s}$\\
SW $\mathrm{H^+}$ density $N_\mathrm{sw} $& 7 $cm^{-3}$\\
SW $\mathrm{H^+}$ temperature $T_\mathrm{p}$ & 15$\times10^4~\mathrm{K}$\\
SW $\mathrm{e^-}$ temperature $T_\mathrm{e}$ & 15$\times10^4~\mathrm{K}$\\
IMF vector $[B_\mathrm{x},\ B_\mathrm{y},\ B_\mathrm{z}]$ & [0, 0, 6]$~\mathrm{nT}$\\
IMF magnitude $\mathrm{B}$ & 6$~\mathrm{nT}$\\
IMF clock angle $\mathrm{\theta}$ & 0$^{\circ}$ \\
Alfv$\acute{e}$n Mach number $M_\mathrm{A}$ & 9.1\\
Magnetosonic Mach number $M_\mathrm{s}$ & 6.7\\
Ion beta $\beta_\mathrm{i}$ & 1 \\
Electron beta $\beta_\mathrm{e}$ & 1 \\
Dipole strength $B_\mathrm{0}$ at the equator on the surface  & 4.5$~\mathrm{\mu T}$\\
\hline
\enddata
\end{deluxetable*}

\begin{deluxetable*}{cccccccccccc}
\tablenum{2}
\tablecaption{A list of parameters for the simulation runs.\label{tab:tab2}}
\tablehead{Runs & $V_\mathrm{sw}$ $\mathrm{(km/s)}$ & $N_\mathrm{sw}$ $\mathrm{(cm^{-3})}$ & $P_\mathrm{d}$ $\mathrm{(nPa)}$ & IMF $B$ $\mathrm{(nT)}$ & Clock angle $\mathrm{\theta}$ ($^\circ$)}
\startdata
1 & -450 & 2.45 & \bf{0.4}  & 6 &0\\
2 & -300 & 7    & \bf{0.5}  & 6 &0\\
\bf{3} & \bf{-450} & \bf{7} & \bf{1.2}  & \bf{6} & \bf{0}\\
4 & -450 & 20   & \bf{3.4}  & 6 &0\\
5 & -800 & 20   & \bf{10.7} & 6 &0\\
\hline
6 & -450 & 7 & 1.2  & 6 &\bf{45}\\
7 & -450 & 7 & 1.2  & 6 &\bf{90}\\
8 & -450 & 7 & 1.2  & 6 &\bf{135}\\
9 & -450 & 7 & 1.2  & 6 &\bf{180}\\
10& -450 & 7 & 1.2  & 6 &\bf{225}\\
11& -450 & 7 & 1.2  & 6 &\bf{270}\\
12& -450 & 7 & 1.2  & 6 &\bf{315}\\
\hline
\enddata
\tablecomments{In Runs 2 and 3 as well as Runs 4 and 5, the solar wind number density is kept constant while the solar wind velocity is increased to raise the dynamic pressure. From Run 3 to Run 4, the velocity remains constant while the density increases. Finally, Runs 1 and 3, with constant velocity and reduced density, produce a low dynamic-pressure case, allowing us to examine whether the results vary monotonically with dynamic pressure.}
\end{deluxetable*}

\clearpage


\begin{figure*}[ht!]
\epsscale{0.95}
\plotone{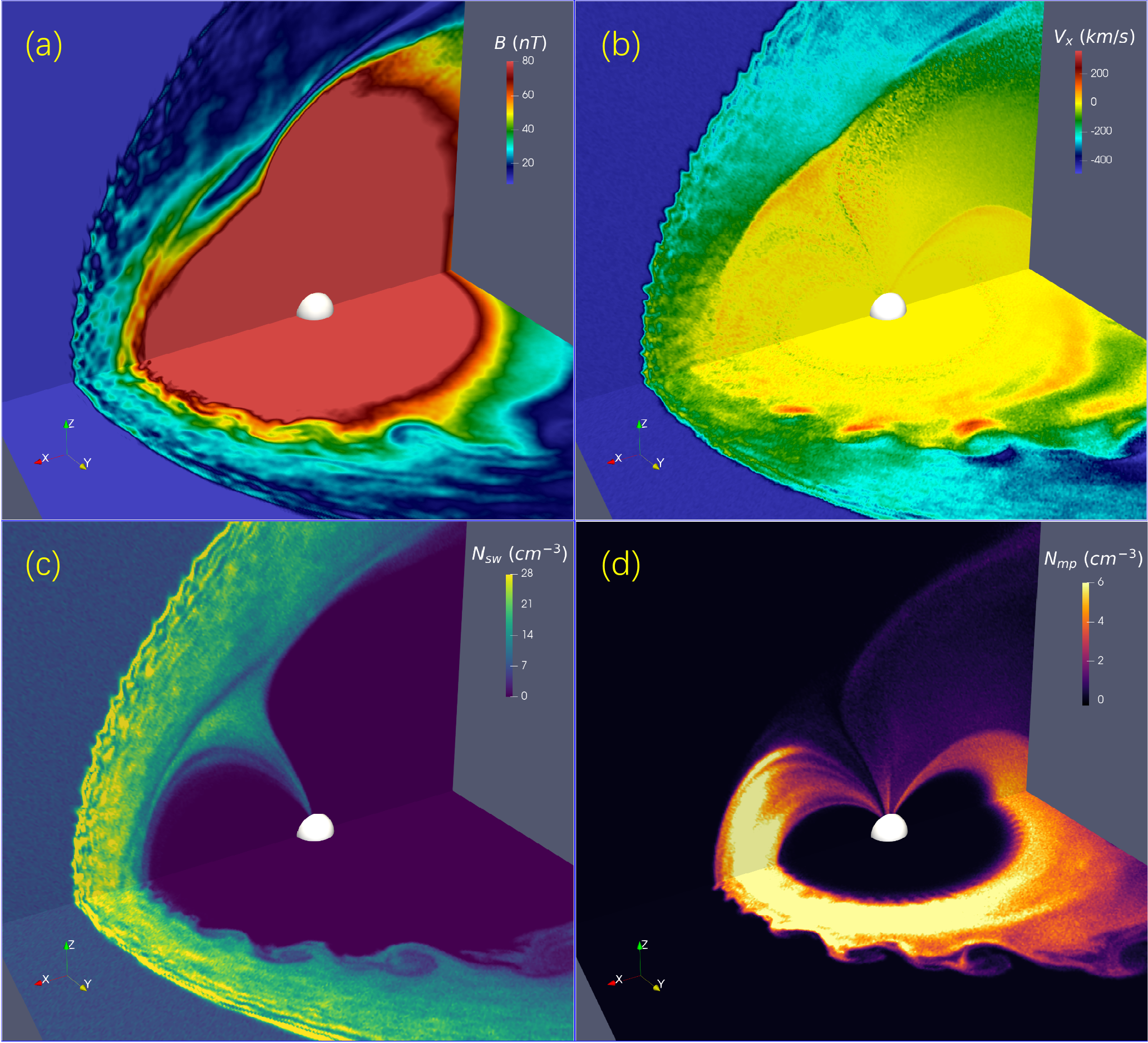}
\caption{\label{fig:fig1}{Magnetopause Kelvin-Helmholtz waves under a northward IMF condition. (a) the magnetic field $\mathrm{B}$ ($\mathrm{nT}$), (b) the ion bulk velocity $V_\mathrm{x}$ ($\mathrm{km/s}$), (c) and (d) number densities of solar wind ions $N_\mathrm{sw}$ ($\mathrm{cm^{-3}}$) and magnetospheric ions $N_\mathrm{mp}$ ($\mathrm{cm^{-3}}$), respectively.}}
\end{figure*}
\FloatBarrier

\begin{figure*}[ht!]
\epsscale{1.15}
\plotone{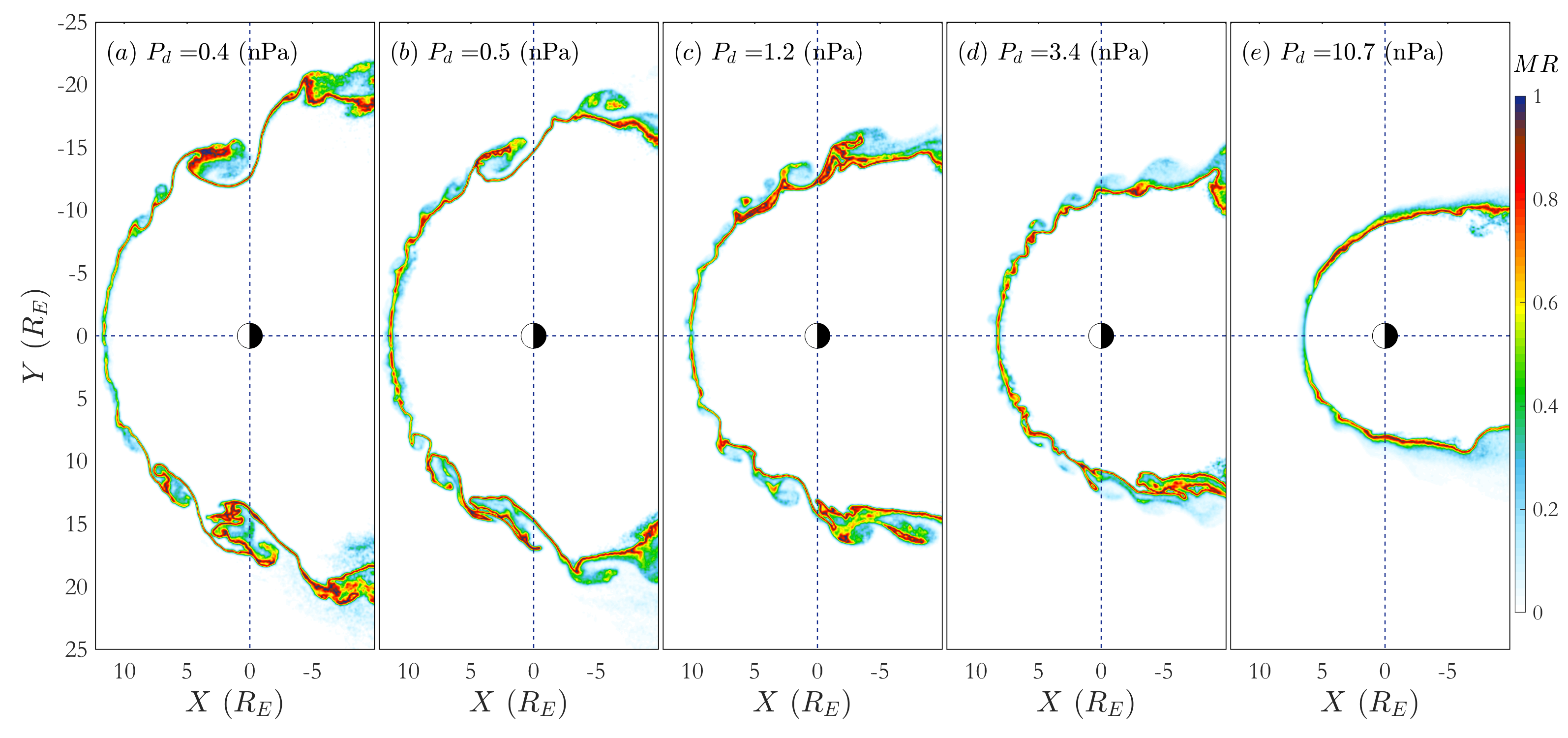}
\caption{\label{fig:fig2}{The influence of solar wind dynamic pressure on ion mixing rates under the same northward IMF condition. The mixing rates are represented in the equatorial plane, where K-H fluctuations are believed to be most intense. From left to right, panels (a-e) demonstrate the trend in mixing rates at the magnetopause boundary layer under conditions corresponding to increasing dynamic pressures ($P$ from $<0.5\ \mathrm{nPa}$ to $>10\ \mathrm{nPa}$).}}
\end{figure*}
\FloatBarrier

\begin{figure*}[ht!]
\epsscale{1.2}
\plotone{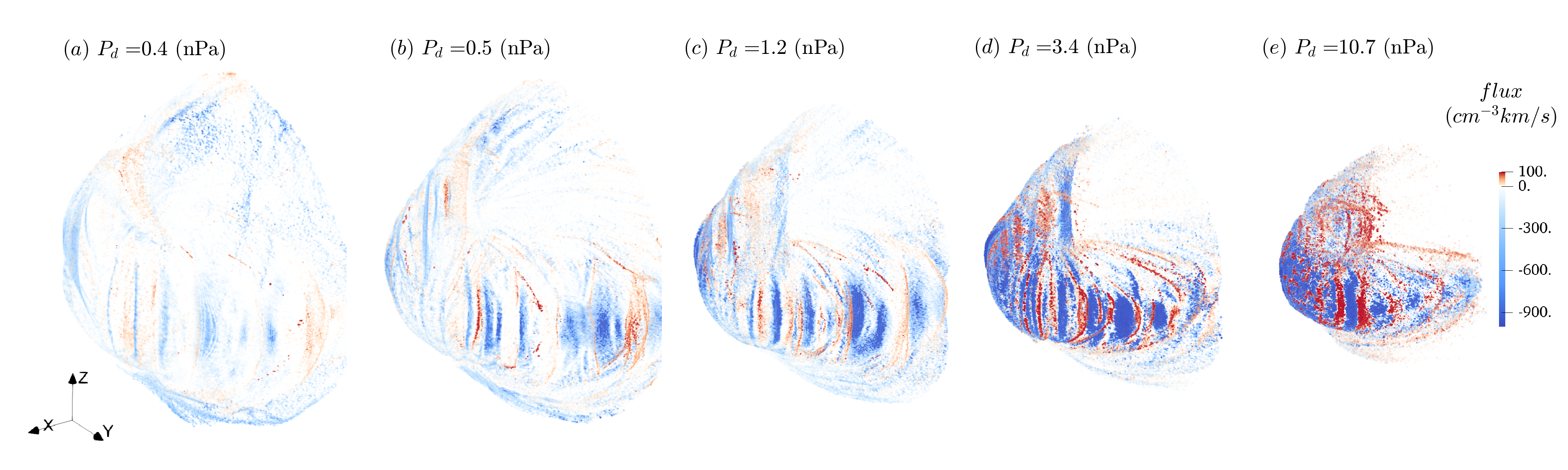}
\caption{\label{fig:fig3}{The corresponding influence of solar wind dynamic pressure on the fluxes of entering solar wind ions (blue) and escaping magnetospheric ions (red) is shown on the magnetopause. The three-dimensional magnetopause surface is obtained from the peaks in mixing rates as depicted in Figure 2.}}
\end{figure*}
\FloatBarrier

\begin{figure*}[ht!]
\epsscale{1.2}
\plotone{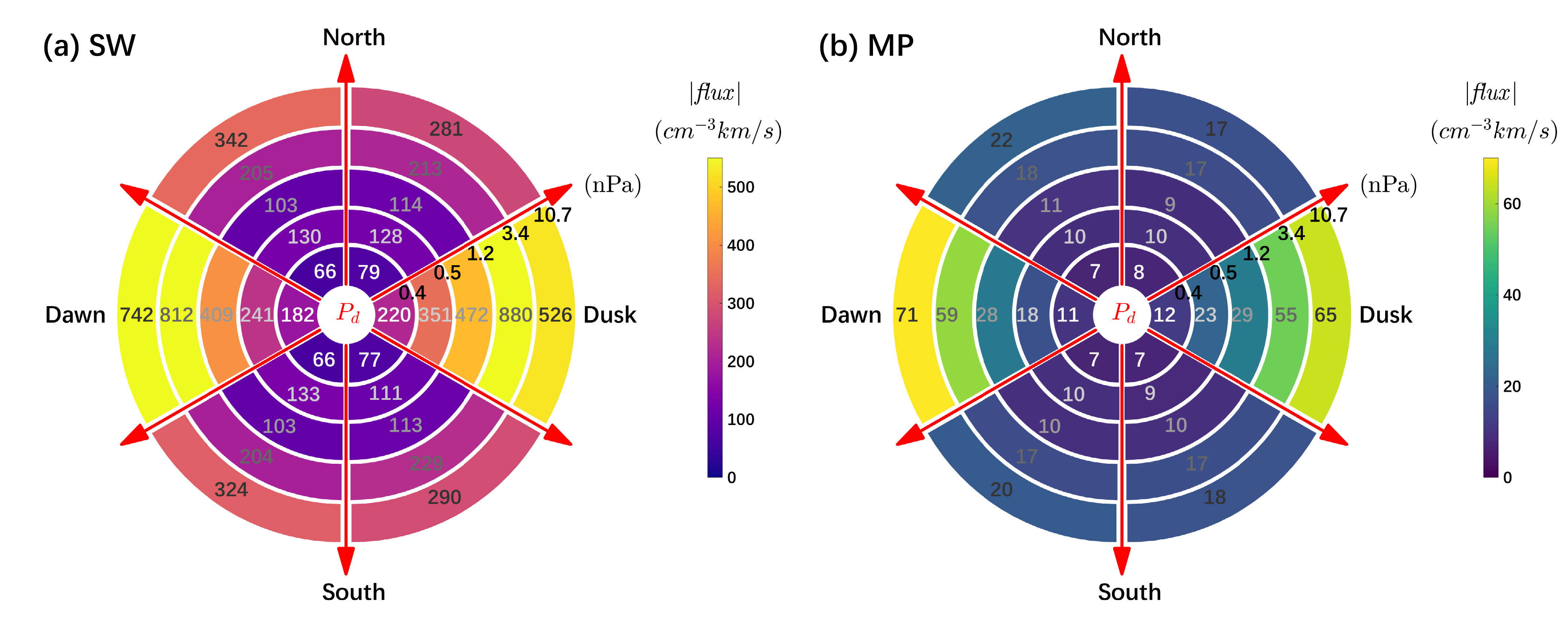}
\caption{\label{fig:fig4}{Quantitative statistical results for the fluxes of entering solar wind ions (a) and escaping magnetospheric ions (b) traversing the magnetopause under different solar wind dynamic pressure conditions. The illustration features five concentric circles, each representing a distinct scenario, arranged from the innermost to the outermost layers, with increasing dynamic pressures ranging from $<0.5~\mathrm{nPa}$ to $>10~\mathrm{nPa}$, as indicated by red arrows. For each concentric circle, average flux measurements in six magnetopause sectors defined by north-south and dawn-dusk orientations are annotated within their respective sectors.}}
\end{figure*}
\FloatBarrier

\begin{figure*}[ht!]
\epsscale{0.98}
\plotone{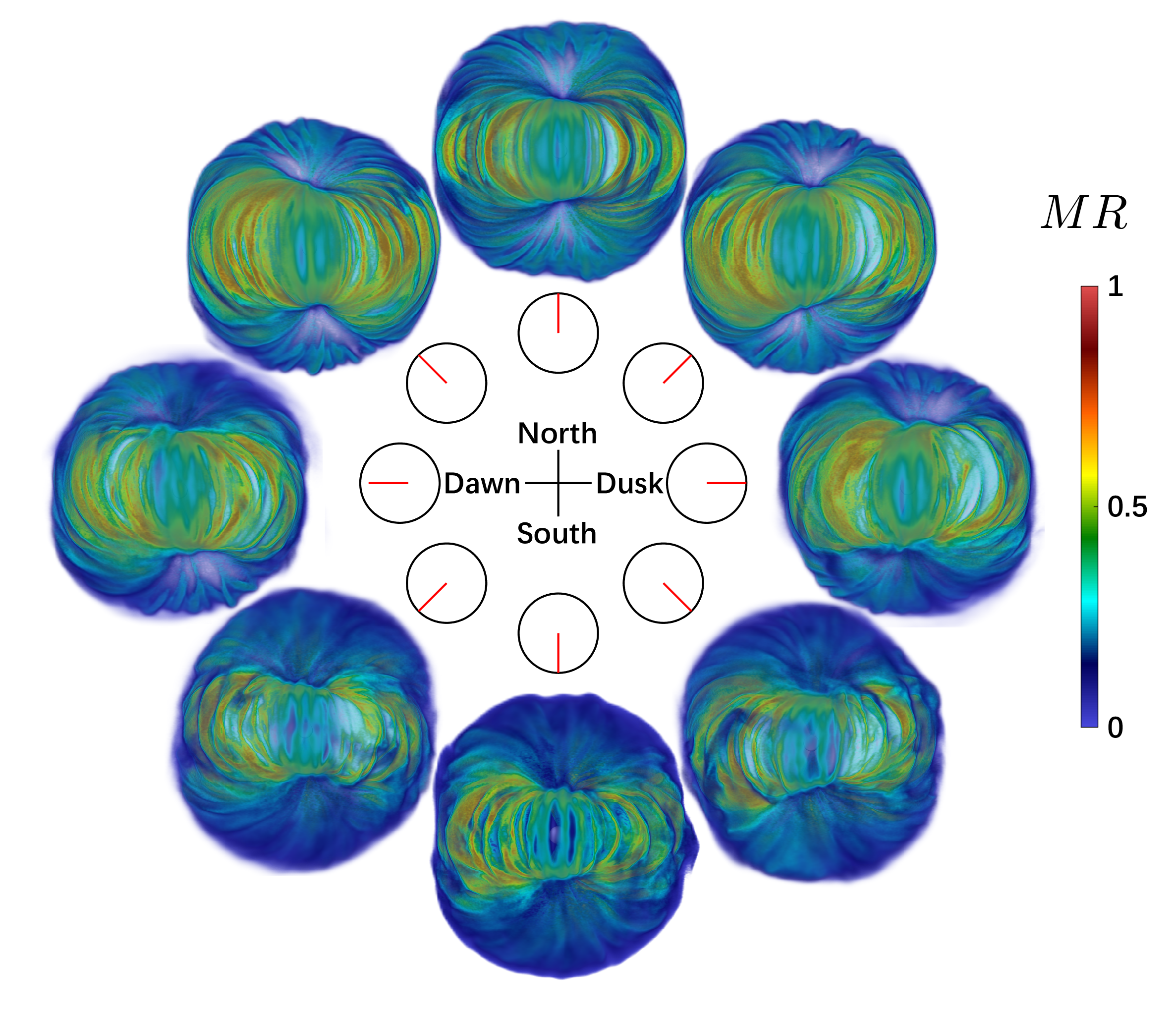}
\caption{\label{fig:fig5}{The impact of IMF clock angle on ion mixing rates. The calculation of mixing rates and the statistics of particle flux are analogous to Figure 2. The perspective of these diagrams is from the Sun looking towards the Earth, with the horizontal axis representing the dawn-dusk direction and the vertical axis indicating geomagnetic north-south poles. Black clocks with red arrows indicate the corresponding IMF clock angle conditions for each case.}}\end{figure*}
\FloatBarrier

\begin{figure*}[ht!]
\epsscale{0.98}
\plotone{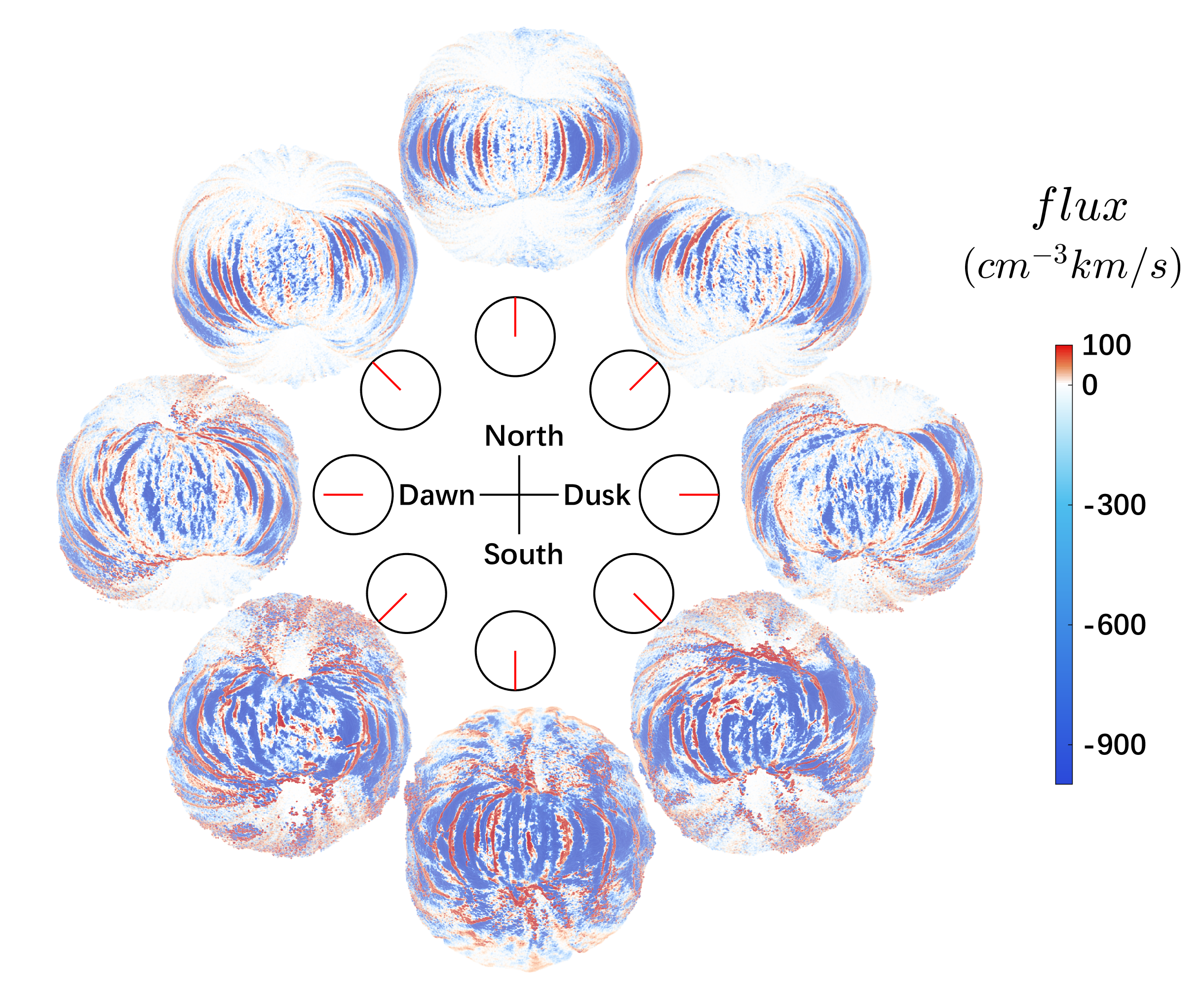}
\caption{\label{fig:fig6}{Corresponding fluxes of entering solar wind ions and escaping magnetospheric ions across the magnetopause under different IMF clock angle conditions.}}
\end{figure*}
\FloatBarrier

\begin{figure*}[ht!]
\epsscale{1.2}
\plotone{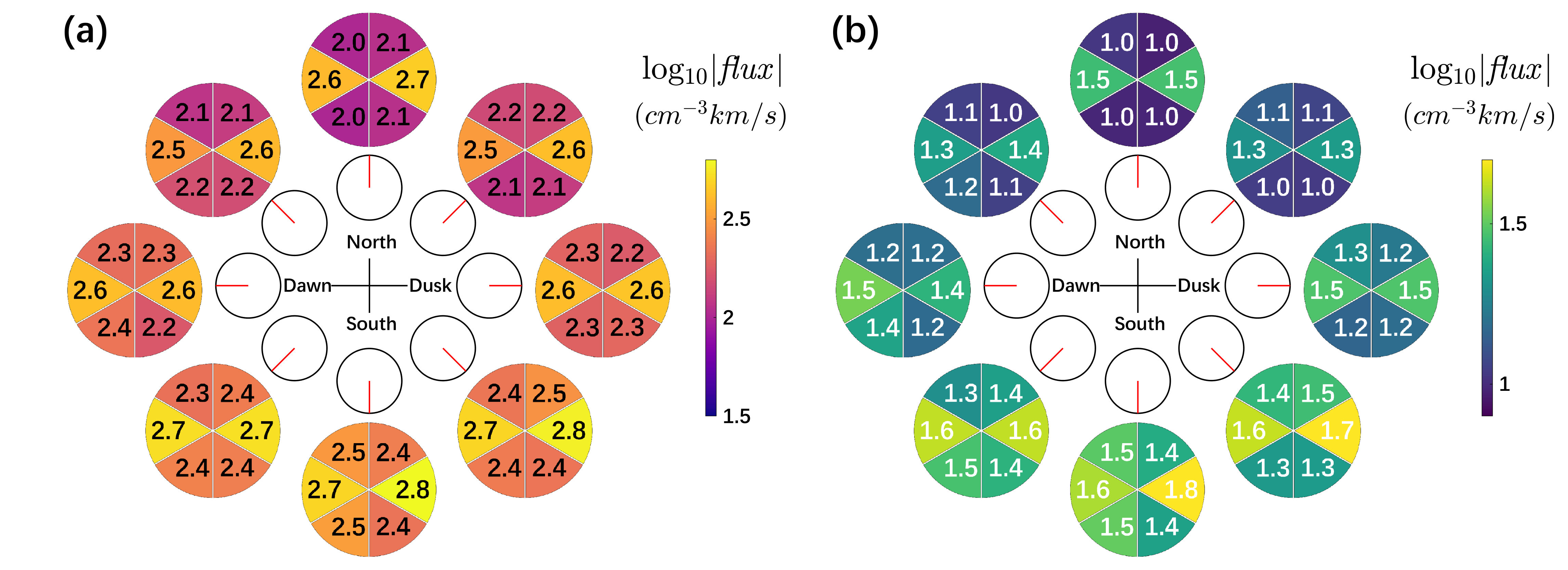}
\caption{\label{fig:fig7}{Quantitative statistical results of the fluxes of entering solar wind ions (a) and escaping magnetospheric ions (b) traversing the magnetopause under different IMF clock angle conditions. In panels (a) and (b), eight colored circles represent eight different clock angle cases. The directions of the clock angles are indicated by red arrows on the adjacent black clocks, respectively. For each case, the magnetopause is divided into six sectors, as illustrated in Figure 4, to quantify and annotate the fluxes in each region.}}
\end{figure*}
\FloatBarrier

\begin{figure*}[ht!]
\epsscale{0.95}
\plotone{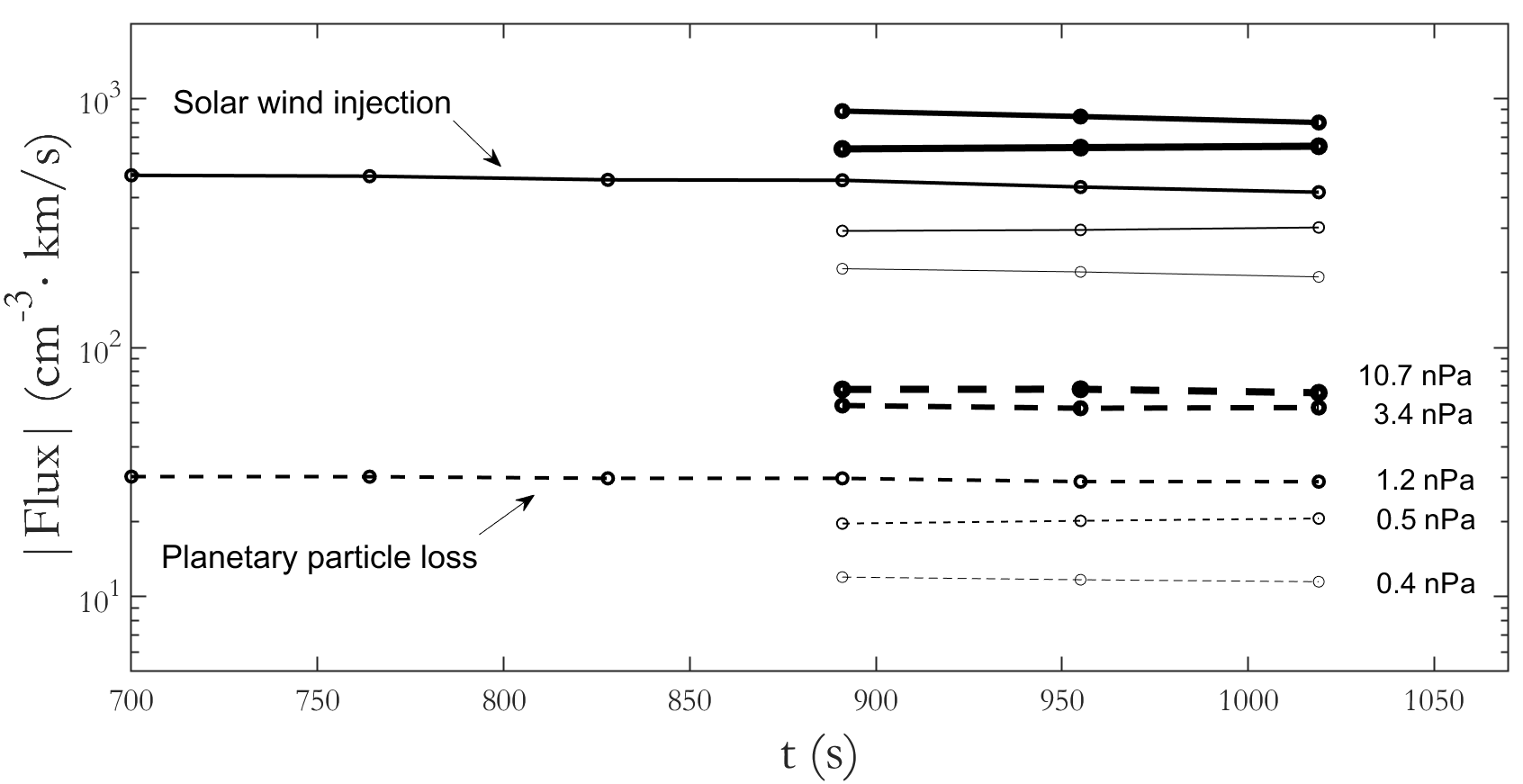}
\caption{\label{fig:fig8}{The time evolution of the fluxes, averaged from the dawn and dusk sectors at low latitudes, for solar wind particle injection (solid) and planetary particle loss (dashed), respectively. The curves, ranging from thin to thick, correspond to results from runs $1\text{--}5$ under varying solar wind dynamic pressure conditions (denoted on the right side of the figure).}}
\end{figure*}
\FloatBarrier
\clearpage


\bibliographystyle{aasjournalv7}
\bibliography{references}{}
%
%

\end{document}